\documentclass[12pt]{article}
\input epsf.sty

\def\lromn#1{\uppercase\expandafter{\romannumeral#1}}

\usepackage{graphicx,amsmath,amssymb}
\begin{document}

\vspace*{1cm}
\begin{center}
\begin{large}
\renewcommand{\thefootnote}{\fnsymbol{footnote}}
\textbf{
Photonic soliton and its relevance to
radiative neutrino pair emission
}\footnote[4]
{Work supported in part by the Grant-in-Aid for Science Research from
the Ministry of 
Education, Science and Culture of 
Japan No. 19204028 and No.19340060}

\end{large}
\end{center}

\vspace{2cm}
\begin{center}
\begin{large}
M. Yoshimura$^{\dagger}$ and N. Sasao$^{\ddagger}$, 

$^{\dagger}$Center of Quantum Universe, Okayama University, \\
Tsushima-naka 3-1-1, Okayama,
700-8530 Japan
\\
$^{\ddagger}$Department of Physics, Kyoto University,\\
Kitashirakawa, Sakyo, Kyoto,
606-8502 Japan 

\end{large}
\end{center}

\vspace{4cm}

\begin{center}
\begin{Large}
{\bf ABSTRACT}

\end{Large}
\end{center}

We consider atomic system of $\Lambda-$type 3-level
coupled to 2 mode fields, and derive an effective Maxwell-Bloch equation
designed for two photon emission between two lower levels.
We find axially symmetric, topologically stable soliton solutions
made of condensed fields.
Immediate implication of soliton formation to radiative neutrino pair
emission is to enhance its rate, larger than
the usual $\propto N^2$ factor  of target number dependence, 
along with another merit of increasing the signal to the background
photon emission.

\newpage
{\bf Introduction}
\hspace{0.5cm}
Superfluorescence(SF) (also called superradiance)
is a phenomenon of cooperative
photon emission triggered by macroscopic
growth of atomic polarization and induced field
\cite{dicke} - \cite{sr review}.
The total emission rate at its maximum scales
with the number of targets $N$  as $\propto N^2$
compared to the spontaneous decay rate $\propto N$.
SF has been observed in a variety of target states,
ranging from gas, solid crystals \cite{sr review}, 
to Bose-Einstein condensates 
\cite{sf in bec 1}, \cite{sf in bec 2}.
A simplest  system of SF may be described by the 
Dicke model \cite{dicke} of 2 levels related by E1 transition.
Its initial state is excited atomic state prepared for instance by
pulsed laser irradiation, with and without 
further triggering field.
SF occurs, in the case of 3 levels, to Raman process, as well
\cite{sf in bec 2}.

The present work has been initiated by our efforts of finding
an enhancement mechanism of radiative neutrino pair emission
from an atomic metastable state
$| 1 \rangle \rightarrow | 2 \rangle + \gamma + \nu_i\nu_j$
($\nu_i$ neutrino mass eigenstates)
\cite{my-06}, \cite{nove08}, in similar ways to SF.
This process is useful to determine all neutrino parameters,
3 masses and 3 mixing angles, and furthermore to
distinguish the Majorana neutrino from the Dirac neutrino.
After the smallest neutrino mass is experimentally determined,
one may proceed to detection of relic neutrino of 1.9 K 
using the Pauli blocking effect \cite{my-taka}.

After derivation of effective Maxwell-Bloch (MB) equation
for related two photon emission process, 
$| 1 \rangle \rightarrow | 2 \rangle + \gamma + \gamma$, 
we realized that non-trivial soliton solutions do
exist, their stability being assured by the topological
winding number associated with SO(2) rotation of
fields.
With formation of solitons
radiative neutrino pair emission is 
enhanced by $O(10^{4}) N^{2}$ 
for targets such as noble gas atoms implanted in solid matrix,
opening a new way to perform neutrino mass spectroscopy,
as envisioned in \cite{nove08}.
Moreover, formation of solitons gives an ideal
mechanism of suppressing the background process
of two photon emission when radiative neutrino pair emission 
is measured.

We shall report on this finding of soliton, 
their fundamental and technological applications, 
which may be wide ranging.
Examples that immediately come in our mind for applications
include the memory storage and quantum computing.

Throughout this work
we assume the natural unit, $\hbar =1$ and $c= 1$.

\vspace{0.5cm}
{\bf Effective Maxwell-Bloch equation}
\hspace{0.5cm}
A standard method of SF analysis is based on the semi-classical
set of Maxwell-Bloch equations \cite{sr review},
involving polarization of atomic targets
and induced electric fields.
We consider a three-level atomic system, $|1 \rangle$ (initial state), $|2 \rangle$ 
(final state) and $|3 \rangle $,
with energy level relation of $\Lambda$ system, $ \epsilon_3 > \epsilon_1 
> \epsilon_2$, $\epsilon_2 = 0$ taken for convenience.
We take into account effects of two fields corresponding to transitions,
$|1 \rangle \leftrightarrow | 3 \rangle $ (pump field ${\cal E}_p$) 
and $| 3 \rangle \leftrightarrow | 2 \rangle$ (Stokes field ${\cal E}_s$).
The interaction Hamiltonian density is 
${\cal H}_I = - (d_1 E R_{31} + d_2 E R_{32} )/2 + {\rm h.c.}$
(we assume two E1 transitions with dipoles $d_i$, 
but extention to M1 transitions
should be evident). 
We may derive MB set of equations from the  equation
for the density matrix, 
$\dot{\rho} = -i[H\,, \rho]$ with $H$ the Hamiltonian, 
by using commutation relations,
$[R_{ij}\,, R_{kl}] = \delta_{jk}R_{il} - \delta_{il}R_{kj}$,
along with the Maxwell equation in medium.
We take the continuum limit so that these transition operators $R_{ij}$
are density functions of time and space coordinates.

When numbers are necessary, we consider two types of examples, 
the case appropriate for
two photon emission such as neutral Ba atom,
and the other case suitable for radiative neutrino
pair emission such as noble gas atoms.
In Ba atom, 
3 relevant states are $| 1 \rangle = ^1$D$_2$,
$| 3 \rangle = ^1$P$_1$ (6s 6p), $| 2 \rangle = ^1$S$_0$.
In this case $d_1 \sim 0.39 \times 10^{-9}$cm,
$d_2 \sim 1.9 \times 10^{-9}$cm, and energy differences are
$\Delta_{31} \sim 0.83$eV, $\Delta_{32} \sim 2.2$eV.
In the case of noble gas atoms 
$| 1\rangle = ^3$P$_2 \,, |3 \rangle = ^3$P$_1 \,, |2 \rangle = ^1$S$_0$
using the LS coupling scheme
(for instance, the more precise configuration for Xe  is
$| 1\rangle = 5p^5(^2P_{3/2})6s ^2[3/2]_2\,, 
|3 \rangle = 5p^5(^2P_{3/2})6s ^2[3/2]_1\,, 
|2 \rangle = 5p^6\, ^1S_0$), 
and Xe example gives $d_2 \sim 2.2 \times 10^{-8}$cm,
$\Delta_{31} \sim 0.12 {\rm eV}\,, \Delta_{32} \sim 8.4{\rm eV}$.
A precise value of $d_1$ for Xe is not known, but we may
infer it of order a typical M1 transition, $O[e/2m_e]$. 

We first make an ansatz for field components $E_i$ 
propagating along z-direction,
\(\:
E_x + iE_y = i\left(
{\cal E}_p^* e^{-i \omega_p (t \pm z)} + 
{\cal E}_s e^{i \omega_s (t \pm z)}
\right) + ({\rm h.c.})
\:\)
and for polarization $R_{ij}$, 
\(\:
R_{31} = R_{31\,, p}e^{-i \omega_p (t \pm z)} 
\,, \;
R_{32} = R_{32\,, s}e^{i \omega_s (t \pm z)}
\,.
\:\)
Unconventional sign mixture $\pm \omega_i \,, i= s\,, p$ in phases here
is chosen for convenience of
discussing two photon process.

The original Bloch equation for the matter system is
\begin{eqnarray}
&&
e^{-i(\Delta_1 - \Delta_2)t}\partial_t R_{21}
= 
 \frac{1}{2} (d_2 {\cal E}_s^*R_{31\,,p} + d_1 {\cal E}_p^* R_{32\,, s}^*)
\,,
\label{start 1}
\\ &&
e^{-i\Delta_1 t}\partial_t (e^{i\Delta_1 t} R_{31\,,p})
=  
- \frac{d_2}{2} e^{-i\epsilon_1 t} 
 {\cal E}_p^* R_{21} + \frac{d_1}{2} {\cal E}_p^* B
- \frac{\kappa_1}{2}R_{31\,,p}
\,,
\label{start 2}
\\ &&
e^{-i\Delta_2 t}\partial_t (e^{i\Delta_2 t}R_{32\,,s})
=  
- \frac{d_1}{2}e^{i\epsilon_1 t}
 {\cal E}_s R_{12} + \frac{d_2}{2} {\cal E}_s C
- \frac{\kappa_2}{2}R_{32\,,s}
\,,
\label{start 3}
\\ &&
\partial_t B = -\Re (2d_1 {\cal E}_p^* R_{31\,,p} 
+ d_2 {\cal E}_s^* R_{32\,,s})
- \frac{2\kappa_1 + \kappa_2}{6}(B+C+n)
\,,
\label{start 4}
\\ &&
\partial_t C = -\Re (d_1 {\cal E}_p^* R_{31\,,p} 
+ 2d_2 {\cal E}_s^* R_{32\,,s})
- \frac{\kappa_1 + 2\kappa_2}{6}(B+C+n)
\,,
\label{start 5}
\end{eqnarray}
and the Maxwell equation in medium,
\(\:
- \partial_t^2 E + \nabla^2 E = \frac{1}{2} \left( 
d_1\partial_t^2 R_{31}
+ d_2 \partial_t^2 R_{32} 
\right) \,,
\:\)
where $B = R_{33} - R_{11}$ and $C = R_{33}- R_{22}$ are 
population difference, $n = n(\vec{x})$ the local number density of target atoms
per unit volume,
and $\Delta_1 = \epsilon_3 - \epsilon_1 + \omega_p$ 
and $\Delta_2 = \epsilon_3 - \omega_s$.
For Raman-like processes $\Delta_i$ are taken as detuning parameters,
and small.
$\kappa_i \propto d_i^2 \Delta_{3i}^3$ are E1 or M1 decay rates 
corresponding to $| 3 \rangle \rightarrow |i \rangle$.

Description of two photon process
$| 1 \rangle \rightarrow | 2 \rangle + \gamma + \gamma$
requires another choice for $\Delta_i$ different from the
Raman process;
$\Delta_1 =\Delta_2  = \Delta 
\sim \Delta_0 \equiv (\epsilon_3 - \epsilon_1 + \epsilon_3 )/2$,
or $\omega_s \sim \omega_p$.
Rapidly oscillating
terms $\propto e^{\pm i \epsilon_1 t}$ 
are averaged out for a long time behavior of variables, 
and one may drop these terms.
We then make slowly varying envelope approximation (SVEA)
by dropping terms $\partial_t R_{3i}$ against $\Delta R_{3i}$, 
which amounts to
balancing equations expressing $R_{3i}\,, i =1\,, 2$ 
in terms of other quantities;
\(\;
 R_{31\,,p} =  (i \Delta + \kappa_1/2)^{-1}d_1B {\cal E}_p^*/2
\,, 
 R_{32\,,s}= (i \Delta + \kappa_2/2)^{-1} d_2C{\cal E}_s/2
\,.
\; \)
MB equation thus derived 
involves an effective direct interaction 
of two photon emission,
$|1 \rangle \rightarrow |2 \rangle$ via pump 
and Stokes field emission;
frequency dependence indeed gives a correct combination 
$\propto {\cal E}_p^*{\cal E}_s^* (B-C)$, with strength $d_1 d_2$.

After a transient time of order the lifetime of
the upper level $|3\rangle$, populations of levels
approach  stationary values, namely time
independent solution of equations for $B\,, C,$
eq.(\ref{start 4}) and eq.(\ref{start 5}), giving
\(\:
B = - n \Delta_{31}^3 |{\cal E}_s|^2/D
\,, \;
C = - n \Delta_{32}^3 |{\cal E}_p|^2/D
\,, \;
D = 
\Delta_{32}^3|{\cal E}_p|^2 + \Delta_{31}^3 |{\cal E}_s|^2
\,,
\:\)
where $d_i |{\cal E}_i| \ll \Delta$ is assumed.
The result implies $B + C = - n$, namely 
$R_{33} = 0 \,({\rm or}\;\ll n)$.

Resulting equations are a closed set for two field amplitudes
and $R_{21}$,
\begin{eqnarray}
&&
\left(\partial_{t}^2 - \partial_{z}^2 - \vec{\nabla}_2^2
\right){\cal E}_s = - \partial_{t}^2 \frac{1}{4\Delta}n d_1^2  
\frac{\Delta_{31}^3 |{\cal E}_s|^2{\cal E}_p^* }
{\Delta_{32}^3|{\cal E}_p|^2 + \Delta_{31}^3 |{\cal E}_s|^2}
\,,
\label{fund eq1}
\\ &&
\left(\partial_{t}^2 - \partial_{z}^2 - \vec{\nabla}_2^2
\right){\cal E}_p = - \partial_{t}^2 \frac{1}{4\Delta}n d_2^2  
\frac{\Delta_{32}^3 |{\cal E}_p|^2{\cal E}_s^* }
{\Delta_{32}^3|{\cal E}_p|^2 + \Delta_{31}^3 |{\cal E}_s|^2}
\,,
\label{fund eq2}
\\ &&
\partial_t R_{21} = i \frac{d_1 d_2 n}{4\Delta}{\cal E}_p^*{\cal E}_s^*
\frac{\Delta_{31}^3 |{\cal E}_s|^2 - \Delta_{32}^3 |{\cal E}_p|^2}
{\Delta_{32}^3|{\cal E}_p|^2 + \Delta_{31}^3 |{\cal E}_s|^2}
\,.
\label{fund eq3}
\end{eqnarray}
The field magnitudes are  limited by eqs.(\ref{start 1}) - (\ref{start 5}),
and not by eqs.(\ref{fund eq1})- (\ref{fund eq2}).
This argument suggests the maximal magnitudes of fields;
$|{\cal E}_s| \leq O[|\partial_t R_{ij}/(R_{ij}d_2 )| ]$ and 
$|{\cal E}_p| \leq O[|\partial_t R_{ij}/(R_{ij}d_1)|]$,
which is later related to the soliton mass ${\cal M}$
by $|\partial_t R_{ij}/R_{ij}| \sim {\cal M}$.

With $\Delta \neq \Delta_0$, these MB equations are also useful for 
description of radiative neutrino pair emission,
$|1 \rangle \rightarrow |2 \rangle + \gamma + \nu_i \nu_j$, 
with a photon energy set at $\epsilon_3 - \Delta$,
when the weak term ${\cal H}_W$ is added to the Hamiltonian density
and treated as a small perturbation.

\vspace{0.5cm}
{\bf Axial symmetry and photonic soliton}
\hspace{0.5cm}
In an axially symmetric case of laser irradiation along z-axis,
we may
introduce cylindrical coordinates, $(z\,, \rho\,, \theta)$.
Spacetime dependence of fields, polarization, population 
difference
\(\:
{\cal E}_j \propto e^{im_j \theta}
\,, \; j = s\,, p \,,
R_{32} \propto  e^{im_s \theta}
\,,
R_{31} \propto  e^{im_p \theta}
\,,
\:\)
is further assumed.
Consistency of angular dependence in equations
(\ref{fund eq1}) and (\ref{fund eq2}) requires $m_s = - m_p = m$.
The requirement of one-valued functions demands that $m$ is an integer.
This introduces the topological winding number $m$.
The transverse operator is then
\(\:
\vec{\nabla}_2^2 = \partial_{\rho}(\rho \partial_{\rho})/\rho
- m^2/\rho^2
\,.
\:\)
We call this topological object the photonic soliton,
in short PS \cite{soliton}.

Fields have polarizations, and we may use this fact to classify
chiralities of solitons.
There are two types of non-trivial topology of
field polarization;
(1) TE mode; this is the case explicitly written above,
$E \sim E_x + iE_y$ such that for instance,
the Stokes field $E_s$ is $  i e^{i{\cal M} (t\pm z) + i m \theta}$ 
times a function dependent on the transverse distance $\rho$.
(2) TM mode; this is the case in which role of the magnetic and the
electric field is interchanged from TE mode.
 $\vec{B} =  \vec{e}_z \times  \vec{E}$
has the similar structure to TE, hence $E_s = - E_y + iE_x$.
Field polarizations are classified  by a set of two opposite numbers 
$(m_s = m\,, m_p = -m)$, the first entry for chirality of
the Stokes field and the second for chirality of the pump.

We further set up an ansatz to work out solutions 
of equations, (\ref{fund eq1}) and (\ref{fund eq2});
a functional form $F(\rho\,, z\,, t) (G_s(\rho)\,, G_p^*(\rho)\,)$
for $({\cal E}_s \,, {\cal E}_p^*)/e^{im \theta}$, with
separation term ${\cal N}^2(\rho) = -(\partial_t^2 F - \partial_z^2 F)/F$
taken more slowly varying with $\rho$ than time variation.
Thus, ${\cal N}^2(\rho) = {\cal M}^2 - \kappa^2 (\rho)$ 
consists of two parts, where
time variation ${\cal M} = i \partial_t {\cal E}_s /{\cal E}_s 
= - i \partial_t {\cal E}_p /{\cal E}_p $
(soliton mass), and variation
along z-direction $\kappa(\rho)$, taken to reflect
effect of index of refraction $\nu$; $\kappa^2 (\rho) = \omega_i^s
+ {\cal M}^2 (\nu^2 - 1)n(\rho)/n_0$, with $n_0$ a central density.
This assumption amounts to a physical picture of
taking field condensates coherently collaborating to
propagate with the same index of refraction.

Using dimensionless quantities,
\(\:
\xi = {\cal M} \rho
\,, \; 
X_m = \sqrt{\xi} G_s/\Delta_{32}^2
\,, \;
Y_{-m} = \sqrt{\xi}G_p/\Delta_{31}^2
\,,
\:\)
and 2-component notation $\psi^T = (X_m\,, Y_{-m})$,
one has 
\begin{eqnarray}
&&
\hspace*{-1cm}
(- \frac{d^2}{d\xi^2}  +  \frac{m^2 - 1/4}{\xi^2} 
- (\nu^2 -1)f(\xi) - \Omega )\psi (\xi) = f(\xi)
\frac{X_m^* Y_{-m}^*}{|X_m|^2 + \eta |Y_{-m}|^2}
\left(
\begin{array}{cc}
\alpha_s \eta^{2} & 0 \\
0 & \alpha_p \eta^{-1}
\end{array}
\right) \psi (\xi)
\,,
\label{xy-eq}
\end{eqnarray}
where 
a diagonal $2 \times 2$ matrix $\Omega$ having
\(\:
\Omega_{11} = \omega_s^2/{\cal M}^2
\,, \; 
\Omega_{22} = \omega_p^2/{\cal M}^2 \,,
\:\)
and
\(\:
\eta = \Delta_{31}/\Delta_{32}
\,, \;
\alpha_s =  d_1^2 n_0/(4 \Delta )
\,, \;
\alpha_p =  d_2^2 n_0/(4 \Delta )
\,, \;
f(\xi) = n(\xi/{\cal M})/n_0
\,.
\:\)
The density profile $f(\xi) = n(\rho)/n_0$ depends on how 
a dense collection of targets (and fields) is excited.
We assume that this profile function has a characteristic length scale
$\rho_0$, which is essentially the soliton size to be determined dynamically.

\vspace{0.5cm}
{\bf Quantum analogy and soliton profile}
\hspace{0.5cm}
We shall study the case
which has the potential of observing radiative neutrino
pair emission;
noble gas atoms (see below on their large rates) implanted with a fraction
$10^{-3}$ in solid para-H$_2$ matrix 
($\rho_0/10 \geq $ the lattice constant of matrix $\sim 3.8 \times 10^{-8}$cm).
The range of parameters are $\alpha_p/\eta = O[0.8({\rm Ne}) - 8({\rm Xe})]
\times 10^{-9}$ and the other $\alpha_s \eta^2$ much smaller,
in the Xe example $\alpha_s \eta^3/\alpha_p = O[10^{-13}]$.

Analogy to quantum scattering problem is useful here.
We first note that effect of the right hand side (RHS) of eq.(\ref{xy-eq})
is described by an effective, non-linear interaction,
\begin{eqnarray}
&&
{\cal H}_{{\rm eff}} = \frac{n}{2\Delta}
\frac{d_1^2 \Delta_{31}^3 |{\cal E}_s|^2 - d_2^2 \Delta_{32}^3 |{\cal E}_p|^2}
{\Delta_{31}^3 |{\cal E}_s|^2 + \Delta_{32}^3 |{\cal E}_p|^2}
\Im ({\cal E}_s^*{\cal E}_p^*) \,,
\end{eqnarray}
giving (non-linear) propagation of Stokes and pump fields,
along with their mixing.
This gives a  Stokes-pump field mixing with effective
strength depending on the field ratio $r = |{\cal E}_p/{\cal E}_s|$
itself;
its effective strength varies from
$n d_1^2/(2\Delta)$
for small $r$ to $- n d_2^2/(2\Delta)$
for large $r$.
Since $d_2 \gg d_1$ usually,
growth of the field ratio is accelerated by
non-linear effect, once it is over a threshold value.
We shall mainly consider the case of small field ratio
(the case of Stokes field dominance), relevant to
radiative neutrino pair emission.
In this case the linearized approximation is excellent, and
two fields are essentially decoupled.

We work out in detail
the linearized approximation for $(X_1\,, Y_{-1})$, using
the density profile $f(\xi) = \xi e^{-\xi/\xi_0}/\xi_0$,
suitable for the use of the Bessel laser beam of order 1 as a trigger,
since it 
gives an interesting scheme of creating fundamental solitons of
chirality $\pm 1$.
The potential 
\(\:
V(\xi) = 3/(4\xi^2) - (\nu^2 - 1) f(\xi)
\:\)
then has repulsion at the origin, and
for a large parameter $(\nu^2 - 1)\xi_0^2$, 
attraction at intermediate region
(and a weak repulsion at infinity).
Since for a large soliton mass ${\cal M}$, the energy
$\Omega_{ii}$ is negligibly small, 
our problem is essentially reduced to finding out (nearly) 
zero energy solutions \cite{resonance},
which exist for discrete set of $\xi_0$, thus forming
an eigenvalue problem for this parameter.

For a crude estimate of the eignevalue $\xi_0$ of zero energy solution,
one may use the WKB formula for energy levels $E_s(\xi_0)$,
namely 
\(\:
2 \int_{\xi_1}^{\xi_2} d\xi \sqrt{E_s - V(\xi\,,\; \xi_0)} = 2\pi \hbar s  \,,
\:\)
($s$ is an integer)
with $\xi_i \,, i=1\,, 2$ turning points,
and set the zero energy condition $E_s(\xi_0)=0$ to derive
eigenvalues of $\xi_0$.
We thus find eigenvalues 
approximately given by $\xi_0 = \xi_s$, with
\(\:
\xi_s = s \sqrt{\pi}/\sqrt{2(\nu^2 - 1)} \,.
\:\)
The number of nodes for the zero energy solution is of order,
$0.8 \sqrt{\nu^2 - 1}\xi_0$.
We illustrate  in Figure 1 a numerical solution of
localized field $(X_1\,, Y_{-1})/\sqrt{\xi} \sim ({\cal E}_s/\Delta_{32}^2\,, {\cal E}_p/\Delta_{31}^2)$.
We confirmed that the WKB energy formula is good for large $s \geq O[10]$.

\begin{figure*}[htbp]
 \begin{center}
 \epsfxsize=0.5\textwidth
 \centerline{\epsfbox{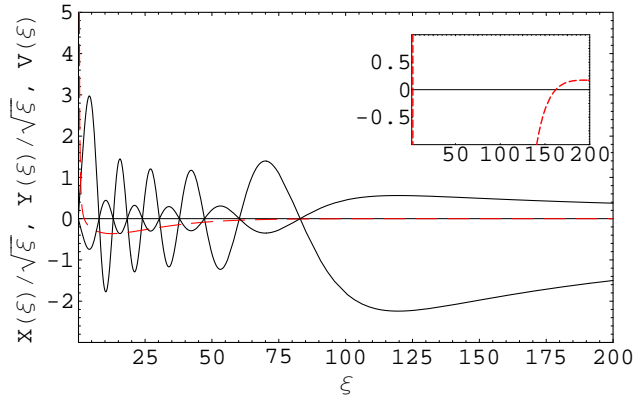}} \hspace*{\fill}
   \caption{Soliton profile. 
   Stokes and pump fields (black) (assumed real)
   and the potential (dashed red) 
   both in arbitrary units
   vs $\xi = {\cal M}\rho$
   are shown for the refractive index
   $\nu=\sqrt{2}$, the field ratio at the origin $ r(0) = - 0.25$
   (essentially fixed at all $\xi$),
    and $\xi_0 = 12.43\cdots $.
   The barrier region of the potential is expanded by $10^4$ in the inset.
   }
   \label{fig:soliton profile}
 \end{center} 
\end{figure*}

Relation between the soliton mass ${\cal M}$ and the soliton size $\rho_0$
is roughly ${\cal M} \rho_0 = O[\xi_s]$, with 
$\xi_s $ one of  the eigenvalues.
The soliton size can be anywhere between atomic distance ($\sim$ 1nm in
solids) and target size, maximally the transverse size of laser
irradiated region, but
most likely sizes of order the wavelength of triggering
laser are the main component.
For numerical estimate below we assume for definiteness
${\cal M} = \xi_s \pi/(2\rho_0)$, often taking $\xi_s = 1$
for crude estimates.

One might characterize the photonic soliton by
saying that it is a stable concentration of
fields, possibly much below the field wavelength scale, supported by surrounding
dense, excited target atoms.

\vspace{0.5cm}
{\bf Implication to radiative neutrino pair emission}
\hspace{0.5cm}
We discuss neutrino pair emission $|1 \rangle \rightarrow
|2 \rangle + \gamma + \nu_i \nu_j$, the pair emission caused by the
Hamiltonian density, ${\cal H}_W = g_W j_{\nu}^{\dagger}R_{31} + ({\rm h.c.})$
with $j_{\nu}^{\dagger}$ the neutrino pair emission current.
Features of a single photon energy spectrum, angular distribution
and distinction from
two photon process have been discussed using a different approach
in \cite{nove08}.
Here we investigate this problem using MB equations,
and examine effect of soliton formation.

The transition amplitude for radiative neutrino pair emission
after soliton formation is  governed by
the following equation written in the interaction picture,
\begin{eqnarray}
&&
\partial_{t} R_{21} = \frac{d_2 g_W }{4\Delta}
n j_{\nu}^{\dagger}{\cal E}_s  \frac{1 - \eta r^2}{1 + \eta r^2}
e^{-i({\cal M} - E_{\nu \nu} - E_{\gamma})(t\pm z)}
\,,
\label{2n eq}
\end{eqnarray}
where $E_{\nu \nu} (E_{\gamma})$ is the energy going into neutrinos
(photons).
A large soliton mass of ${\cal M} \gg  \Delta $ appears
in the exponential factor.

The large time limit for the transition rate per unit volume
$d|R_{21}(t)|^2/dt$  can be worked out
by using the wave function of soliton 
${\cal E}_s(\rho)  e^{im \theta}
e^{\pm i({\cal M} - E_{\nu \nu} - E_{\gamma})z}$, 
resulting in a similar
formula to the Fermi golden rule, with a difference
of large energy factor ${\cal M} - E_{\nu \nu} - E_{\gamma}$.
How the total rate scales with the total number of target atoms
is as follows.
We may take the maximal magnitude 
${\cal E}_s$ of order ${\cal M}/d_2 $,
along with $r = 0$.
The  enhancement factor per unit atom is 
\(\:
K = 
|{\cal E}_s/e|^2 (V \Delta^3/2\pi^2)
(2\Delta/{\cal M}) \sim 1/(\pi d_2^2 \rho_0 \Delta^3)\,,
\:\)
where $ e \sim \sqrt{\Delta/2V} \sim \Delta^2/\sqrt{2}$ is the field
magnitude of a single atom transition, and the inverse of the last
factor ${\cal M}/2\Delta$ is the number of events for the soliton
decay. One has the entire enhancement factor for $N$ target atoms
given by $K N^2$, where
\begin{eqnarray}
&&
K \sim 2.4 \times 10^8 \xi_s
(\frac{{\rm nm}}{\rho_0})
(\frac{{\rm eV}}{\Delta})^3 (\frac{10^{-8}{\rm cm}}{d_2})^2
\,.
\end{eqnarray}

What happens is formation of soliton of 
mass ${\cal M}$, which macroscopically decays.
This massive soliton goes into many radiative pairs.
The microscopic description of this phenomenon
is that the rate of elementary radiative neutrino
pair emission is enhanced by an extra factor, 
$O[{\cal M}/(d_2^2 \Delta^3) \sim 1/(d_2^2 \rho_0 \Delta^3) ]$
in addition to the usual coherence factor $N^2$.

The elementary rate $\gamma$
of radiative neutrino pair emission is roughly given by \\
\( \:
\gamma_{\gamma \nu \nu} = G_F^2 \Delta_{12}^5 (\gamma_{32}/\Delta_{32}) 
(\Delta_{12}/\Delta_{31})^2/(15\pi^5)
\sim 3.3 \times 10^{-34} s^{-1}
(\Delta_{12}/{\rm eV})^5
(\gamma_{32}/\Delta_{32})
(\Delta_{12}/\Delta_{31})^2 \,.
\:\)
For noble gas atoms implanted with a fraction $10^{-3}$ 
in solid para-H$_2$ matrix, the maximum rate is
$4 \times 10^{-27}s^{-1} N^2 $ for Ar and 
$2 \times 10^{-28}s^{-1} N^2 $ for Xe,
giving the total rate of order (40 - 2)Hz 
for $N = 10^{14}$
(we took $\xi_s = 1$ for this estimate).  
Other noble gas atoms in solid matrix give
similar rates, somewhat larger for Ne by $O[400 ]$ than Xe.
Alkaline earth and other atoms often give much smaller rates. 
The enhanced rate scales with 
$\Delta_{12}^4 \Delta_{32}^2/(\Delta_{31}^2 \rho_0)$ of target parameters,
which works to give large rates for the $\Lambda-$type noble gas atoms,
with large $\Delta_{i2}$ and small $\Delta_{31}$.

The precise angular distribution of photon in radiative neutrino
pair emission depends on how the triggering laser irradiation
leads to formation of solitons, their number and their size distribution,
which is a difficult problem to solve.
But, the angular distribution 
from decay of a single soliton can be worked out
from (\ref{2n eq}).
Without much calculation we may deduce basic features
of photon angular distribution 
by noting combined spatial variation of field and
neutrino pair ${\cal E}_s j_{\nu}^{\dagger}$. The phase factor in the
exponent, which needs to be canceled, is 
$ \left ( (\pm {\cal N} + K_{\rho})\rho
+  (\pm \kappa  + K_z) z + (m + m_{\nu \nu})\theta \right) $,
with $K_i$ the momenta of many  neutrino pairs.
The correlation to the cylinder axis is evident, and
the photon emission is confined to a small angle region
of $\theta \leq O[\kappa/{\cal M} = (\nu^2 - 1)n(\xi_0)/n_0 ]$.

On the other hand, two photon process has in RHS of
eq.(\ref{2n eq})
\(\:
d_1 d_2  n {\cal E}_s^* {\cal E}_p^*
(1 - \eta r^2) /(4\Delta (1 + \eta r^2)\,)e^{iE_{\gamma \gamma}(t\pm z)}
\,,
\:\)
in which the soliton mass ${\cal M}$ is missing in the exponent.
Thus, two photon emission from solitons do not occur.
The two photon process however can occur from amplified pump and
Stokes field not related to soliton formation, 
to give a rate simply proportional to $N^2$.
Thus, the rate for radiative neutrino pair emission,
is more enhanced, at least by the factor 
$r^{-2}{\cal M}/(d_2^2 \Delta^3) \sim r^{-2}/(d_2^2 \rho_0 \Delta^3)$, 
than for two photon emission, which is of order $10^5 r^{-2}$ or more
for noble gas atoms in solid matrix.
Incidentally, the elementary rate for two photon
emission from a single atom
is estimated $O[1] {\rm sec}^{-1}$ for noble gas atoms.

Controled two photon emission, however, becomes possible by using a
systematic destruction of coherence, such as abrupt modulation
of dielectric constant.
It should also be noted that
stability, and possibility of controled coherence breaking, of PS
gives an ideal mechanism of enhancing the signal to the 2 photon 
background ratio in measurement of forbidden processes.

The potential background of multi-photon (more than 2 photon) emission 
is not enhanced
at all by soliton formation, thus when the elementary rate of multi-photon
QED process is smaller than the enhanced rate of radiative neutrino pair
emission, say $O[1]$Hz, the multi-photon emission does not become the major
background.

\vspace{0.5cm}
{\bf Applications and outlook}
\hspace{0.5cm}
Here we briefly discuss some possible technological
applications using two photon emission caused by
controled destruction of stable PS's.

Topological solitons, both stable or unstable (the case of resonance), 
are likely to be
created in the region of high dielectric constant $\epsilon$.
One may use for preparation the
photonic crystal type of medium \cite{photonic crystals}
doped by target atoms
of long-lived $\Lambda-$type level such as Ba D-levels.
Many photonic solitons of small size may be created when an array disk  
made of rectangular shaped high $\epsilon$ material
is irradiated by the Bessel laser beam of chirality 1
for the trigger. This might
serve for the memory storage.
On the other hand, when a cylinder made of many high $\epsilon$ tubes
is irradiated by a Bessel beam of large aperture, 
one may expect creation of many PS's of long size, which might serve
for efficient light transportation.
From energetic reasons we expect that
PS's of size of order the laser wavelength are
more likely to be created at its formation.

Correlation of emitted two pulsed lights
after controled coherence breaking is excellent,
in direction, energy, and chirality.
Correlated emission of strong light pulses after
PS destruction may thus be useful for quantum information.

What is pressing is experimental confirmation
of the basic idea in the present work, and is to clarify how
easy or how difficult it is to creat many PS's
using material technologically available at present.
On theoretical side calculation of dynamical time
evolution is left to further work.

\vspace{0.5cm}
{\bf Acknowledgements}
\hspace{0.5cm}
We should like to thank for our collaborators
of SPAN group, A. Fukumi, K. Nakajima, I. Nakano,
and H. Nanjo for enlightening discussions on 
this and related subjects.


\begin{thebibliography}{99}
\bibitem{soliton}
To the best of our knowledge,
the terminology of photonic soliton
has been used in the literature in rahter
imprecise ways.
Our terminology is based on the topological
winding number and solutions of MB equation,
and has no direct connection to the ones used in literature.
\bibitem{resonance}
For non-zero $\Omega_{ii} > 0$
there may exist, for a large
$\sqrt{\nu^2-1}\xi_0$, resonance solutions in their energy vicinity.
They are
time dependent solutions of the form,
$\psi (x\,, t) = e^{- iE_* t - \gamma_* t/2}u(x)$, where
\(\:
\left( - d^2/dx^2 + V(x) \right) u(x) = (E_* - i \gamma_*/2)u(x)
\:\)
with resonance parameters, real $E_*\,, \gamma_* > 0$.
Their presence implies instability when we consider time evolution
towards soliton formation.
Unstable resonances, if their lifetimes are large enough, 
are however useful to initiate two photon emission.
\end{thebibliography}
\end{document}